\documentclass{emulateapj}
\newcommand{\beq}{\begin{equation}}
\newcommand{\eeq}{\end{equation}}

\def\AMd{$d_\mathrm{AM\,CVn} = 606^{+135}_{-93}$~pc}
\def\HPd{$d_\mathrm{HP\,Lib} = 197^{+14}_{-12}$~pc}
\def\CRd{$d_\mathrm{CR\,Boo} = 337^{+44}_{-35}$~pc}
\def\VCd{$d_\mathrm{V803\,Cen} = 347^{+32}_{-27}$~pc}
\def\GPd{$d_\mathrm{GP\,Com} = 75^{+2}_{-2}$~pc}

\def\fs{\hbox{$.\!\!^{\rm s}$}}

\def\farcs{\hbox{$.\!\!^{\prime\prime}$}}
\def\gsim{\mathrel{\hbox{\rlap{\lower.55ex \hbox {$\sim$}}
                   \kern-.3em \raise.4ex \hbox{$>$}}}}
\def\lsim{\mathrel{\hbox{\rlap{\lower.55ex \hbox {$\sim$}}
                   \kern-.3em \raise.4ex \hbox{$<$}}}}
\newcommand{\gpm}[3]{$#1^{+#2}_{-#3}$}
\newcommand{\tnm}[1]{\tablenotemark{#1}}

\begin{document}

\shorttitle{HST/FGS Parallaxes of AM CVn Stars}
\shortauthors{G.\,H.\,A. Roelofs et al.}

\title{HST/FGS Parallaxes of AM CVn Stars and Astrophysical Consequences\footnote{Based on 
observations made with
the NASA/ESA Hubble Space Telescope, obtained at the Space Telescope
Science Institute, which is operated by the
Association of Universities for Research in Astronomy, Inc., under NASA
contract NAS 5-26555}} 

\author{G.\,H.\,A.~Roelofs\altaffilmark{1}, P.\,J.~Groot\altaffilmark{1},
G.\,F.~Benedict\altaffilmark{2}, B.\,E.~McArthur\altaffilmark{2},
D.~Steeghs\altaffilmark{3,4}, L.~Morales-Rueda\altaffilmark{1},
T.\,R.~Marsh\altaffilmark{4}, and G.~Nelemans\altaffilmark{1}}

\altaffiltext{1}{Department of Astrophysics, Radboud University Nijmegen, Toernooiveld 1, 6525 ED, Nijmegen, The Netherlands}
\altaffiltext{2}{McDonald Observatory, University of Texas, Austin, TX 78712}
\altaffiltext{3}{Harvard--Smithsonian Center for Astrophysics, 60 Garden Street, Cambridge, MA 02138}
\altaffiltext{4}{Department of Astrophysics, University of Warwick, Coventry CV4 7AL, UK}

\email{g.roelofs@astro.ru.nl}

\begin{abstract}
We present absolute parallaxes and relative proper motions for five AM
CVn stars, which we obtained using the Fine Guidance Sensors on board the {\it Hubble Space Telescope}.
Our parallax measurements translate into distances \AMd; \HPd; \CRd; \VCd; and \GPd.
From these distances we estimate the space density of AM CVn stars and suggest that previous estimates have been too high by about an order of magnitude. We also infer the mass accretion rates which allows us to constrain the masses of the donor stars, and we show that relatively massive, semi-degenerate donor stars are favored in all systems except GP Com.
Finally, we give updated estimates for their gravitational-wave signals, relevant for
future space missions such as the proposed \emph{Laser Interferometer Space Antenna (LISA)}, based on their distances and the inferred masses of the binary components. We show that all systems but GP Com are excellent candidates for detection with \emph{LISA}.
\end{abstract}

\section{Introduction}
The AM CVn stars are white dwarfs (WDs) accreting matter from a degenerate or semi-degenerate companion, constituted primarily of helium with traces of metals (but no hydrogen). Because of their evolved nature they exist in ultra-compact configurations, with orbital periods ranging from about one hour down to ten, possibly even five, minutes. General overviews of this class of ultra-compact binary stars are given by \citet{War95} and, more recently, \citet{Nel05}.

The formation of AM CVn stars is as yet poorly understood, although several formation channels have been proposed to contribute significantly to the AM CVn population. The first is stable Roche-lobe overflow in a (formerly detached) double white dwarf binary, when the two white dwarfs are brought together by angular momentum loss due to the emission of gravitational waves (the WD channel; see \citealt{Nel01}). The second is Roche-lobe overflow from a helium star onto a white dwarf, followed by a period minimum around ten minutes, caused by the quenching of helium fusion and the donor star becoming semi-degenerate -- equivalent to the orbital period minimum in the hydrogen-rich Cataclysmic Variables (CVs) (the helium-star channel; e.g. \citealt{Ibe91,Nel01}). In a third scenario, a hydrogen-main-sequence star starts mass transfer onto a white dwarf right at the time when its core is becoming depleted of hydrogen, shifting the normal CV orbital period minimum down to, again, about ten minutes depending on the level of hydrogen depletion (the evolved-CV channel; \citealt{Pod03}).

Determining which of these evolutionary channels actually produce AM CVn stars, and in what numbers, has been a long-standing problem. An important observable is the distance to known AM CVn stars: this helps in determining their space density, and their absolute magnitudes contain important information about the rate at which they accrete matter, which in turn constrains the masses of the donor stars.

In this paper we present \emph{Hubble Space Telescope (HST)} parallaxes of five AM CVn stars, ranging from 17 to 46 minutes in orbital period. They were selected on their average $V$-band magnitude ($V<16$) because of the source brightness limitations of the Fine Guidance Sensors (FGS). All five systems show quite characteristic accretion disk behavior. The shortest-orbital-period systems AM CVn and HP Lib, at 17 and 18 minutes respectively, have an apparently stable disk that appears optically thick with shallow helium absorption lines (e.g.\ \citealt{Gre57,ODo94}), while the longest-period system GP Com, at 46 minutes, has a stable disk that appears optically thin with strong helium emission lines (e.g.\ \citealt{Nat81}). The two intermediate systems under consideration, CR Boo and V803 Cen at orbital periods of 24 and 27 minutes, display transitions between high and low accretion disk states, with the disk changing from absorption-line to emission-line states and the luminosity dropping by as much as five magnitudes (e.g.\ \citealt{Woo87,ODo87}). This is presumably caused by a thermal instability in their disks \citep{Tsu97}. Photometrically, these AM CVn stars thus behave much like the well-studied CVs (e.g.\ \citealt{Pat97,Pat00}).

The outline of this paper is as follows. We describe the observations and data reduction in Section \ref{obssec}, and present the parallaxes and absolute magnitudes in convenient tabular form in Section \ref{ressec}. We then discuss the implications of these results for the space density of AM CVn stars, for their mass accretion rates and the masses of the components, and for their gravitational-wave signals in Section \ref{dissec}.

\section{Observations and Data Reduction}
\label{obssec}
Ten sets of astrometric data were acquired with the Fine Guidance Sensors on board \emph{HST} for each of
our five science targets. At each epoch we measured
several reference stars and the target multiple times to correct for
intra-orbit drift. We obtained these sets in pairs typically
separated by a week. Each complete data aggregate spans
$\sim$1.5 years, except for GP Com, where a servicing mission forced a
one-year slip of an observation at one maximum parallax
factor. Table~\ref{tbl-LOO} contains the epochs of observation and
measured photometry for each AM CVn star. The data were
reduced and calibrated as detailed in Benedict et al.\ (2002a, 2002b, 2005)\nocite{Ben02a,Ben02b,Ben05}, \citet{McA01}, and \citet{Sod05}.

Because the parallaxes determined for the AM CVn stars have been
measured with respect to reference frame stars which have their own
parallaxes, we had to either apply a statistically derived correction
from relative to absolute parallax \citep{WvA95} or estimate the absolute parallaxes of the reference
frame stars (listed in Table \ref{tbl-POS}). We chose the second method, as it yields a more direct (less Galaxy-model dependent) way of determining the reference star absolute parallaxes.

In principle, the colors,
spectral type, and luminosity class of a star can be used to estimate
the absolute magnitude, $M_V$, and $V$-band absorption, $A_V$. The
absolute parallax is then simply,
\beq
\pi_\mathrm{abs} = 10^{-(V-M_V+5-A_V)/5}
\eeq

To obtain the spectral type and luminosity class of all reference stars, we combined existing photometric data from the Two-Micron All Sky Survey (2MASS), proper motions from the USNO CCD Astrograph Catalog (UCAC2), and ground-based spectroscopic follow-up from the 2.5-m Isaac Newton Telescope (with IDS), the 4.2-m William Herschel Telescope (with ISIS), the 6.5-m Magellan--Baade telescope (with IMACS), and the FLWO 1.5-m (with FAST), where the acronyms in parentheses are the names of the spectrographs used. The spectra typically had a $\sim$5\,\AA\ resolution and a wide spectral range, suitable for classification purposes from early- to late-type stars.
Table \ref{tbl-SPP} lists the spectral types and luminosity classes we obtained for our reference stars based on their independent photometric and spectroscopic classifications. Estimated classification uncertainties were used to
obtain the errors on the $m-M$ values in that table.

Assuming an $R=3.1$ Galactic reddening law (Savage \& Mathis
1979\nocite{Sav79}), we derived $A_V$ values by comparing the measured
colors (Table \ref{tbl-IR}) with intrinsic $(V-K)_0$ colors from
Bessell \& Brett~(1988)\nocite{Bes88} and \citet{Cox00}. The resulting $A_V$ values are collected in Table \ref{tbl-AV}, from which we calculated a field-wide average $A_V$ to be used in Equation 1. The resulting reference star parallax estimations are listed in Table \ref{tbl-SPP}.

\section{Results}
\label{ressec}
\subsection{Parallaxes and Absolute Magnitudes}
\label{parsec}
Table \ref{partable} lists the absolute parallaxes, proper motions, and absolute magnitudes we obtained for the five AM CVn stars.
To determine the average absolute magnitudes, we first derived average values for the apparent magnitudes of our objects. AM CVn, HP Lib and GP Com have never been observed to show brightness variations much larger than about a tenth of a magnitude around their mean, and it is commonly assumed that their accretion disks are in stable states of high (AM CVn, HP Lib) or low (GP Com) mass transfer, such that the thermal instabilities that may trigger transitions between different brightness states do not occur. We could therefore use our accurate FGS photometry (Table \ref{tbl-LOO}), which consists of 10 visits per object well separated in time, to directly derive average apparent magnitudes for these three systems. The two other objects, V803 Cen and CR Boo, are known to show complex brightness variations, as also evidenced by our FGS photometry (Table \ref{tbl-LOO}). For these we used the results of extensive photometric monitoring campaigns conducted by Patterson et al.\ (1997, 2000)\nocite{Pat97,Pat00} to estimate time-averaged apparent magnitudes, by simply averaging the flux levels reported for these systems over the multiple years of observations. For all systems, we have made the crucial assumption that the average flux levels at which they have been observed since their discovery are representative for their longer-term average flux levels. The results are collected in Table \ref{partable}.

When using a trigonometric parallax to derive absolute magnitudes for a class of objects, a correction is usually made for the Lutz--Kelker (LK) bias \citep{Lut73}.
This LK bias occurs when the number of stars in a population significantly increases (or decreases) with distance around the measured distance for a certain object in the population, both due to an increase in the sampled volume with distance and a decrease in the population space density with distance. This can be corrected for if one knows the scale height of the population under consideration. For the AM CVn stars, the scale height is not known and has to be estimated.

GP Com, together with its twin V396 Hya, quite likely belongs to a (potentially large) halo population of old and very metal-poor AM CVn stars \citep{Mar99, Mor03}. For the shorter-period systems, we have no indication that they belong to a halo population. Significant amounts of metals have been detected in AM CVn, HP Lib and V803 Cen (Roelofs et al.\ 2006b,c\nocite{Roe06b,Roe06c}), suggesting a thin or possibly thick disk origin. Similar data is as yet unavailable for CR Boo, but several more recently discovered AM CVn stars -- SDSS J124058.03$-$015919.2 \citep{Roe05}, and V406 Hya \citep{Roe06a} -- show abundances of metals in optical spectra (typically Fe, Mg and Si) that are compatible with solar values. There is thus no evidence that the AM CVn stars in the solar neighborhood in general belong to a halo population. In addition, none of our objects show particularly large transverse velocities perpendicular to the Galactic Plane (see Table \ref{partable} for their proper motions). We thus assume that AM CVn, HP Lib, CR Boo and V803 Cen all belong to a disk population.

We model the space density of this population with a thin disk component of moderate age, with a scale height $h_{z,\mathrm{thin}}=300$\,pc, plus an older thick disk component of scale height $h_{z,\mathrm{thick}}=1250$\,pc that contributes 2\% at $z=0$\,pc, where $z$ is the distance from the Galactic Plane:
\beq
\frac{\rho(z)}{\rho(0)} = 0.98\,\,\mathrm{sech}\left(z/300\,\mathrm{pc}\right) + 0.02\,\,\mathrm{sech}\left(z/1250\,\mathrm{pc}\right)
\label{popmodel}
\eeq

We then calculate the LK bias $\Delta\pi$ numerically, as this allows for a relatively easy derivation of the LK bias correction (see below on how we apply this) for an arbitrarily complicated distribution of stars $\rho(\pi)$ by evaluating
\beq
\Delta \pi(\mathrm{LK}) = \pi - \frac{
\int\pi'\mathrm{d}\pi'\,\frac{\rho(\pi')}{\pi'^4}\,\exp\left(-\frac{(\pi'-\pi)^2}{2\sigma^2_\pi}\right)
}
{
\int\mathrm{d}\pi'\,\frac{\rho(\pi')}{\pi'^4}\,\exp\left(-\frac{(\pi'-\pi)^2}{2\sigma^2_\pi}\right)
}
\eeq
where $\rho(\pi)$ will in general depend on the coordinates (but in our model population, only on the Galactic latitude) of the object under study, and the $\pi^{-4}\mathrm{d}\pi$ term is simply the increase in (phase) space with $\pi$. We perform this convolution integral between $\pi\pm5\sigma_\pi$; for parallax measurements with $\sigma_\pi/\pi\geq 0.2$, an LK bias correction will become increasingly unrealistic (e.g.\ \citealt{Lut73}), although in our case the decreasing $\rho(\pi)$ with $\pi\rightarrow 0$ due to the finite population scale height keeps the integral well-behaved. The fact that our sample is, essentially, magnitude limited, causes an additional suppression of the population density $\rho(\pi)$ towards $\pi\rightarrow 0$ (e.g.\ \citealt{Smi03}; also \citealt{Lut73}). However, the object for which this potentially matters due to its relatively large measurement error (i.e.\ AM CVn; $\sigma_\pi/\pi=0.18$) is at high Galactic latitude such that the assumed finite scale height has already effectively killed off the population $\rho(\pi)$ at small $\pi$.

GP Com, for which we assumed a halo origin, essentially represents the limiting case $\rho(\pi)\rightarrow\mathrm{constant}$ since its observed distance is much smaller than any reasonable choice of scale height for the halo.

The resulting LK bias corrections on the absolute magnitudes, $\Delta M(\mathrm{LK})$, are given in Table~\ref{partable}. We have used them to convert the inferred absolute magnitude for e.g.\ AM CVn, based on its measured distance, to the average absolute magnitude that \emph{a sample of objects exactly like AM CVn} would have:
\beq
M_V = \langle V\rangle - \left(m-M\right) - A_V + \Delta M(\mathrm{LK})
\eeq
That is, we apply the LK bias correction to the $M_V$'s of the individual objects in our sample. In practice, these $\Delta M(\mathrm{LK})$'s are actually quite small compared to the `raw' measurement uncertainties in the distance moduli $m-M$. The errors on these corrections, also given in Table \ref{partable}, have been estimated by varying the modeled thin disk scale height from 200--500\,pc.
All results, including the final absolute magnitudes we obtain for the five stars, are listed in Table~\ref{partable}.

\begin{deluxetable*}{rccccc}
\tablecaption{AM CVn Parallaxes, Proper Motions, and Absolute Magnitudes\label{partable}} 
\tablewidth{0pt}
\tablehead{
\colhead{Parameter} & \colhead{AM CVn} &\colhead{HP Lib}&\colhead{CR Boo}&\colhead{V803 Cen}&\colhead{GP Com}
}
\startdata
Study Duration (yr)                             &1.59           &1.53           &1.62           &1.51           &2.4\\
Observation Sets (\#)                           &10             &10             &9              &10             &10\\
Reference stars (\#)                            &3              &5              &5              &5              &5\\
Reference stars $\langle V\rangle$              &13.98          &13.18          &12.09          &14.58          &14.50\\
Reference stars $\langle B-V\rangle$            &0.84           &0.75           &0.81           &0.67           &0.80\\[0.7ex]

{\it HST} $\pi_\mathrm{abs}$ (mas)              &1.65$\pm$0.30  &5.07$\pm$0.33  &2.97$\pm$0.34  &2.88$\pm$0.24  &13.34$\pm$0.33\\
{\it HST} relative $\mu$ (mas yr$^{-1}$)        &34.25$\pm$0.88 &33.59$\pm$1.54 &38.80$\pm$1.78 &9.94$\pm$2.98  &352.36$\pm$12.79\\
in Position Angle (\arcdeg)                     &67.0$\pm$1.7   &314$\pm$14     &$-79.9\pm3.7$  &248$\pm$11     &$-84.7\pm3.1$\\[0.7ex]

$d$~(pc)                        &\gpm{606}{135}{93}     &\gpm{197}{14}{12}      &\gpm{337}{44}{35}      &\gpm{347}{32}{27}      &\gpm{75}{2}{2}\\[0.7ex]
$\langle V\rangle$                              &14.02$\pm$0.05 &13.59$\pm$0.05 &14.5$\pm$0.2   &14.0$\pm$0.2   &15.94$\pm$0.05\\
$A_V$                                           &0.05$\pm$0.02  &0.34$\pm$0.06  &0.03$\pm$0.04  &0.31$\pm$0.09  &0.02$\pm$0.11\\[0.7ex]
$m-M$                           &\gpm{8.91}{0.44}{0.36} &\gpm{6.47}{0.15}{0.14} &\gpm{7.64}{0.26}{0.24} &\gpm{7.70}{0.19}{0.17} &\gpm{4.37}{0.05}{0.05}\\[0.7ex]
LK bias $\Delta M$                              &$-0.16\pm0.05$ &$-0.03\pm0.00$ &$-0.09\pm0.01$ &$-0.06\pm0.00$ &$-0.01\pm0.00$\\[0.7ex]
$M_V$                           &\gpm{4.90}{0.37}{0.45} &\gpm{6.75}{0.16}{0.17} &\gpm{6.74}{0.32}{0.33} &\gpm{5.93}{0.28}{0.29} &\gpm{11.54}{0.13}{0.13}\\
\enddata
\end{deluxetable*}

\section{Discussion}
\label{dissec}
\subsection{Previous Distance Measurements}
\label{sec:comparisonprev}
\subsubsection{Ground-based parallax measurements}
Two of our targets (AM CVn and GP Com), as well as one other member of the AM CVn class (V396 Hya) have
a measured ground-based parallax. \citet{Tho03} derived a
parallax for GP Com of 14.8$\pm$1.3 mas,
translating into a distance of \gpm{68}{7}{6} pc, in good
agreement with our determination. For V396 Hya, J.~R. Thorstensen
(private communication) derives a parallax of 12.9$\pm$1.4 mas, translating into
a distance of \gpm{76}{11}{8} pc, very close to the distance found for
GP Com.   

A ground-based absolute parallax of $4.25\pm0.43$ mas is derived for AM CVn itself (C. Dahn 2004, as quoted by \citealt{Nel04}), translating into a distance of 235 pc. This is in disagreement with our measurement at the 5-$\sigma$ level. The origin of this rather significant discrepancy is not clear. We have no reason to question the FGS measurements, but we do note that there were only three suitable reference stars in the field of AM CVn, whereas we used five for the other sources. This makes AM CVn's measurement more susceptible to errors in the reference star grid. On the other hand, if AM CVn really were at 235\,pc, it seems highly unlikely that we should not have detected a much larger signal.

None of the other AM CVn stars have a ground-based parallax
determination sofar. A lower limit to the distance to CR Boo is set by
C. Dahn (2004, as quoted by \citealt{Esp05}) of $d_{\rm
CR\,Boo}>250$ pc, fully consistent with our measurement. 

\subsubsection{Model-based distances}
A number of accretion disk modeling studies have been done, where the observed optical spectra, in particular the shapes of the spectral lines and the overall slope of the spectrum, are fitted to accretion disk models to derive a number of parameters, including the distance. Estimates include: $d_{\rm ES\,Cet} = 350$ pc \citep{Esp05};
$d_{\rm CR\,Boo} = 469\pm50$ pc, $d_{\rm HP\,Lib}=188\pm50$ pc, $d_{\rm V803\,Cen}=405\pm50$ pc, $d_{\rm
AM\,CVn}=288\pm50$ pc (all from \citealt{Nas01}); and
$d_{\rm AM\,CVn}=420\pm80$ pc, $d_{\rm CR\,Boo}=206\pm15$ pc
\citep{Elk00}.

Comparing these values with our parallaxes, the large discrepancy with
the distance to AM CVn is again striking, whereas the values for CR
Boo scatter around our parallax value, and the determination for HP
Lib is dead-on target. The different model predictions for AM CVn and CR Boo differ quite a bit, the difference being formally significant at the 5-$\sigma$ level for CR Boo.

Based on our \emph{HST/FGS} parallaxes, it would thus appear that ground-based parallaxes up to at least 75
pc can give quite reliable results, whereas ground-based parallaxes at several hundred pc as well as distances obtained from disk modeling remain uncertain.

\subsection{Space Density of AM CVn Stars}
\label{densec}
Based on an estimated $M_V=9.5$ for the AM CVn stars in their high states, and the resulting distances $d<100$\,pc for five of the six AM CVn stars known at the time, \citet{War95} derived a local space density $\rho(0)\sim3\times10^{-6}$\,pc$^{-3}$ for the systems that are (mostly) in a high state.

Our \emph{HST} parallaxes show that the absolute magnitude in the high states is close to $M_V=6$, significantly brighter than the value used by \citet{War95}. Despite the increase in the number of known AM CVn stars in recent years, mainly due to the Sloan Digital Sky Survey \citep{And05,Roe05}, the number of known systems that have high states of $m_V \leq 14.5$ is still four. It would thus seem that we can lower the estimates by \citet{War95} that were based on this sample. We employ the ``$1/V_\mathrm{max}$'' method of \citet{Sch75}, generalized to include the variation of space density with Galactic height:
\beq
\rho(0) = \frac{1}{4\pi\beta} \sum_\mathrm{obj} \frac{1}{\int_0^{r_\mathrm{max}}\! r^2 \mathrm{d}r\, \rho(z)/\rho(0)}
\eeq
Here $r_\mathrm{max}$ is the maximum distance at which an object would still just belong to the sample at $m_V=14.5$ given its $M_V$, and $0<\beta\leq 1$ is an (unknown) completeness parameter that corrects for the systems that would have belonged to our sample had they been discovered. The decrease in space density with Galactic height $z=r\sin b$, i.e.\ the term $\rho(z)/\rho(0)$, is modeled as in Section \ref{parsec} (Eq.\ (\ref{popmodel})). Summing over the four known objects in our sample then leads to a local space density of $\rho(0)=2\times10^{-8}\beta^{-1}$ pc$^{-3}$ for the short-period AM CVns that have high states. Assuming a sample completeness $\beta$ of only 10\%, based on the notion that there are still large parts of the sky such as the Galactic Plane that are poorly surveyed, our result is still an order of magnitude lower than Warner's (1995) \nocite{War95} estimate for this population.

If we combine the above result with the notion that AM CVn stars with orbital periods $P_\mathrm{orb}\lesssim2000$\,s are observed to be in a high state at least some of the time, and we compare with evolutionary predictions that only about 2\% of the local AM CVn population should currently be at these or shorter orbital periods \citep{Nel01,Nel04}, we arrive at a local space density estimate
\beq
\rho(0)\approx1\times10^{-6}\beta^{-1}\,\,\mathrm{pc}^{-3}
\eeq
for the entire AM CVn population, including the old and dim ones at long orbital periods.

If we again allow for a sample completeness $\beta$ of only 10\%, our space density for the entire population is still an order of magnitude lower than predictions from population synthesis models by \citet{Nel01,Nel04}, which indicate $\rho(0)\sim 1\times 10^{-4}$\,pc$^{-3}$ (there is a small dependency of our `observed' space density on the population models, because the 2\% fraction of systems at $P_\mathrm{orb}<2000$\,s was derived from them). It thus appears that space density estimates -- both of the bright, short-period ones and of the entire population -- have so far been too high by about an order of magnitude. The sample completeness estimate $\beta$ is nevertheless still very uncertain, and a more homogeneous sample of AM CVn stars will be needed to bring down the corresponding uncertainty in the space density \citep{Roe07}.

\subsection{Mass Accretion Rates}
\label{accsec}
From the absolute magnitudes we can estimate the
mass accretion rates, $\dot M$, in our AM CVn stars, assuming that the observed flux is dominated by the accretion luminosity. For this we need to estimate
what fraction of the total accretion luminosity of the stars is emitted in the
$V$ band; that is, we need the bolometric correction.

\begin{figure*}
\begin{center}
\plotone{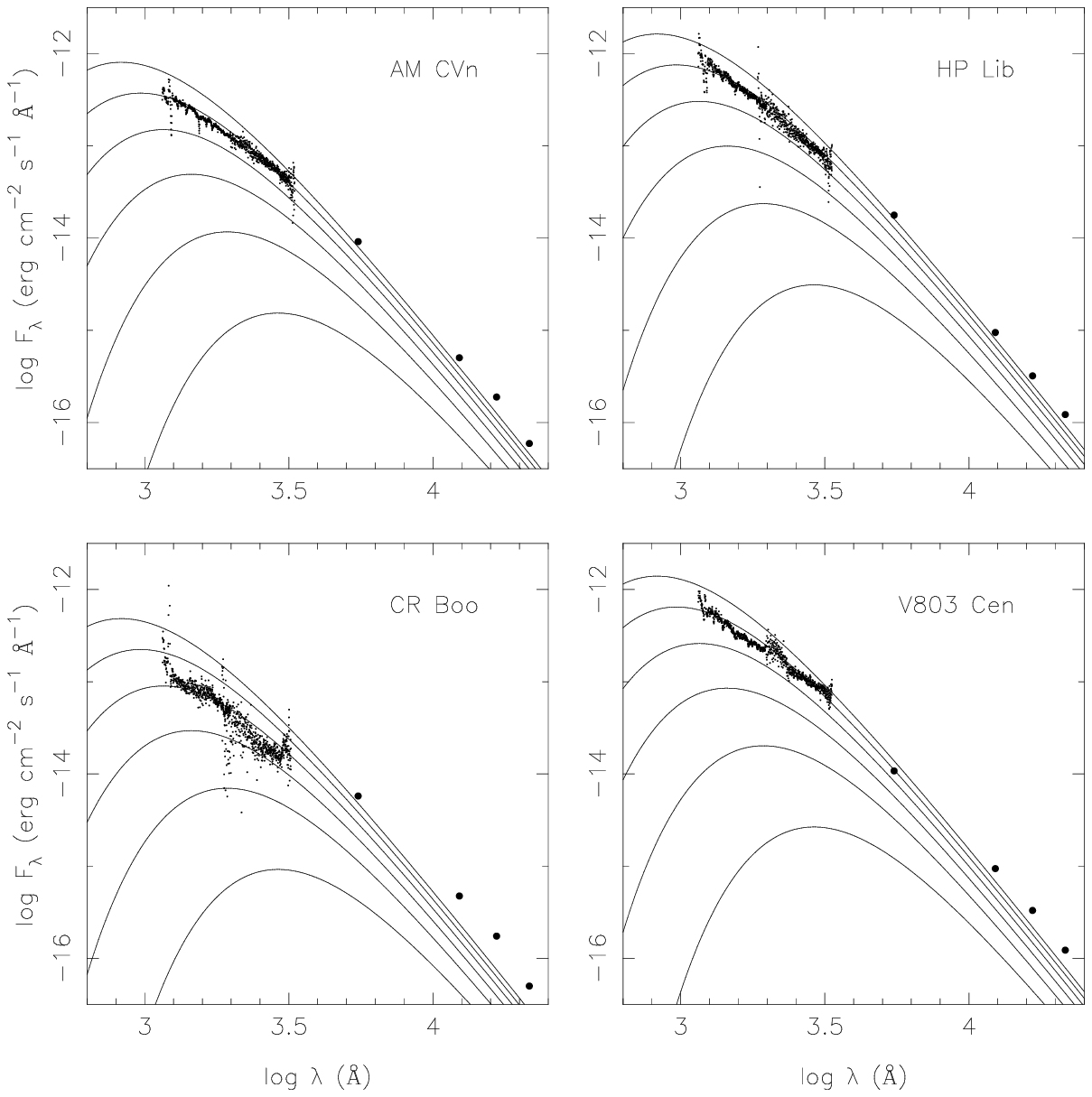}
\caption{Observed spectral energy distributions of AM CVn, HP Lib, CR Boo, and V803 Cen, compiled from archival \emph{IUE} spectra (small dots), the optical V-band fluxes from table \ref{partable} (left-most large dot), and the near-infrared J-, H-, and K-band fluxes from 2MASS (three right-most dots). The solid lines are blackbody spectra of $T_\mathrm{eff}=10-35$\,kK in steps of 5\,kK (bottom to top) to get an idea of the temperatures. For CR Boo and V803 Cen, we selected the apparent high-state spectra from the \emph{IUE} archives; however, we cannot exclude that they represent slightly lower intermediate states. For CR Boo this was definately the case for its short-wavelength (1150--2000\,\AA) spectrum; we scaled it up to match the overlapping long-wavelength (1850--3350\,\AA) \emph{IUE} spectrum. Its UV flux still seems low with respect to the average optical flux and the 2MASS snapshots. The data have been corrected for extinction according to \citet{Car89} and assuming an $R_V=3.1$ extinction law.}
\label{sedfig}
\end{center}
\end{figure*}

To determine this correction, we compose the spectral energy distributions (SEDs) for all systems except GP Com, based on the optical fluxes, archival \emph{IUE} spectra covering the far- and near-UV, and their 2MASS detections. See Figure~\ref{sedfig}. The most reliable SEDs are those of the non-outbursting systems AM CVn and HP Lib; that of AM CVn has the additional advantage that the extinction towards AM CVn is low thanks to its high Galactic latitude, reducing the impact of uncertainties in the extinction correction. We see that the slopes of the UV spectra suggest at least $\sim$30\,kK `blackbody' components, although both in the infrared and at the far-UV end of the spectra there is a flux excess relative to a 30\,kK blackbody, indicating that the systems are not described perfectly by a single-temperature blackbody. At the far-UV end, we could be seeing a contribution from the accreting white dwarf, which is expected to have an effective temperature of $\sim$40\,kK \citep{Bil06}. However, this contribution to the UV flux has to be small since the quiescent magnitudes of V803 Cen and CR Boo indicate that the accretor contributes no more than a few per cent of the light in the optical, while the accretor is not expected to be much hotter than the disk. The donor star is expected to have an effective temperature of only a few thousand Kelvin (C. Deloye, private communication; \citealt{Del07}), while it should not be larger than the disk; therefore it should not contribute significantly to the SED either. Thus we are most likely seeing a disk with a range of temperatures.

We first derive a \emph{minimum} bolometric correction (BC) for AM CVn by summing all the \emph{observed} fluxes, without extrapolating beyond the observed wavelengths. This yields a lower limit of $\mathrm{BC}\leq-2.2$. We then have to estimate how much additional flux is beyond the observed wavelengths. By far the largest part of this flux will be at wavelengths shorter than the observed range. From the slopes of the far-UV spectra, which do not appear to be flattening off that much towards shorter wavelengths, we estimate that they continue as 30\,kK blackbodies. With this estimated extra flux the total bolometric correction becomes $\mathrm{BC}=-2.5\pm0.3$, where the error has been estimated by considering a 40\,kK instead of a 30\,kK blackbody. Given the similarity of the SEDs, we assume this bolometric correction for all systems (although CR Boo's \emph{IUE} spectrum has a lower UV flux compared to its average optical flux, we assume that this is due to its outburst behavior).

The exception is GP Com, which is thought to be in a stable state of low mass transfer. Its SED has been shown to match rather well with a $T_\mathrm{eff}=11$\,kK blackbody, which is expected to be the accreting white dwarf \citep{Bil06}, plus helium emission lines from the accretion disk which appears optically thin in the continuum. We have therefore used a bolometric correction
appropriate for an $11\pm1$\,kK blackbody: BC$_{\rm GP\,Com}=-0.5\pm 0.2$. Since the temperature of the accreting white dwarf we see today is probably set by accretion heating that occurred a long time ago, when GP Com was a much shorter-period binary with much higher $\dot M$ \citep{Bil06}, we can only put an upper limit to GP Com's present-day accretion rate.

For a given system with a given BC, we can derive the bolometric luminosity, $L$, and compare with the value
\beq
L = \frac{1}{2} \dot{M_2} \big(\Phi(L_1) - \Phi(R_1)\big)
\label{Lbol}
\eeq
expected for conservative mass transfer through a Keplerian (i.e.\ virialised) disk, where $\Phi$ is the Roche potential, at the inner Lagrange point $L_1$ and the surface of the accreting star $R_1$. Since the accreting white dwarf is expected to be rather hot due to accretion-induced heating \citep{Bil06}, at least for the four bright systems, we use accretor radii that are 5--10\% (depending on mass) larger than the idealized zero-temperature radii, based on the mass--radius relations given by \citet{Pan00}. These correspond to a core temperature of about $3\times 10^7$\,K in the models of \citet{Bil06,Bil07}.

We can then link the mass transfer rate to the component masses by assuming that gravitational-wave losses drive the evolution of the system, as is commonly assumed \citep[e.g.][]{Del05, Mar04}, and requiring that the secondary fill its Roche lobe. This gives an equilibrium mass transfer rate
\beq
\frac{\dot{M_2}}{M_2} = \frac{\dot J}{J} \frac{2}{\zeta_2+5/3-2q}
\eeq
where the angular momentum loss due to the emission of gravitational waves is given by
\beq
\frac{\dot J}{J} = -\frac{32}{5}\frac{G^3}{c^5} \frac{M_1 M_2 (M_1+M_2)}{a^4}
\label{Jdot}
\eeq
\citep{Lan71}, $q\equiv M_2/M_1$ is the mass ratio of the binary, $a$ is the separation between the donor mass $M_2$ and the accretor mass $M_1$, and
\beq
\zeta_2 \equiv \frac{\mathrm{d} \log R_2}{\mathrm{d} \log M_2}.
\eeq
expresses the mass--radius relation of the mass-losing star. The latter may range from $\zeta_2\approx -0.06$ for a semi-degenerate helium object based on evolutionary calculations by \citet{Tut89} as parameterized by \citet{Nel01}, to $\zeta_2\approx-1/3$ for an (idealized) zero-temperature white dwarf.

For a given mass ratio $q$ and a choice of mass--radius relation for the secondary star $\zeta_2$, the mass of the secondary $M_2$ is now fixed by the observed bolometric luminosity, if one makes the crucial assumption that the present-day accretion rate corresponds to the binary's equilibrium rate for gravitational-wave-driven mass transfer. At present only GP Com (Steeghs et al.\ in preparation) and AM CVn \citep{Roe06b} have a directly measured mass ratio; for the other three we can either estimate the primary's mass, or the mass ratio. In \citet{Roe06b} we derive a primary mass $M_1\approx 0.7 M_\odot$ for AM CVn, and combined with the mass distribution of single white dwarfs (e.g.\ \citealt{Ber92}), it would seem reasonable to adopt a similar canonical white dwarf mass for the accretors in HP Lib, CR Boo, and V803 Cen.

We do, however, have one additional piece of information on these three systems: they all show `superhump' periods $(P_\mathrm{sh})$ in time-series photometry (e.g.\ \citealt{Pat97,Pat00,Pat02}), which are thought to be due to a tidal resonance between the accretion disk and the secondary star. Observationally and theoretically, it has been found that there exists a (roughly linear) relation between the superhump period excess $\epsilon$, defined as
\beq
\epsilon \equiv \frac{P_\mathrm{sh} - P_\mathrm{orb}}{P_\mathrm{orb}}
\eeq
and the mass ratio $q$, at least for the hydrogen-rich Cataclysmic Variables \citep{Pat05,Whi88,Whi91,Hir90}. To date, there is only one AM CVn star for which we have both a kinematically derived mass ratio and a superhump period excess, namely AM CVn itself \citep{Ski99, Roe06b}. We may estimate mass ratios for HP Lib, CR Boo, and V803 Cen by assuming a linear $\epsilon(q)$ relation,
\beq
\epsilon(q) = 0.12 q
\label{epsilon}
\eeq
for them, based on $q=0.18$, $\epsilon=0.0218$ measured for AM CVn. The resulting mass ratios for HP Lib, CR Boo and V803 Cen are shown in Table~\ref{mdottable}. The first two of these are about twice as large as those used by \citet{Del05}, who employ an $\epsilon(q)$ relation that is based on fits to numerical simulations of accretion disks in hydrogen-rich CVs, as given by \citet{War95}. Adopting instead the empirical $\epsilon(q)$ for hydrogen-rich CVs from \citet{Pat05} already yields larger mass ratios than the ones used in \citet{Del05}, but still smaller than ours. Our mass ratio estimate for V803 Cen is also much larger \emph{relative} to the other systems than the one used by \citet{Del05}; this is due to a new robust measurement of V803 Cen's orbital period, as will be discussed in section \ref{qV803Cen}. Although our $\epsilon(q)$ relation is uncertain since it is based solely on AM CVn (see also \citealt{Pea07} for a recent discussion), it is probably the best we can do until independent mass ratio measurements become available for more AM CVn stars. We choose to employ this $\epsilon(q)$ relation combined with the known superhump excesses for HP Lib, CR Boo, and V803 Cen to estimate their individual mass ratios, rather than assume a canonical primary mass for all three.

Lastly, a correction has to be made for the inclination $i$ of the binary, since the `apparent absolute magnitude' \citep{War95} depends on the inclination. We thus link the observed absolute magnitude $M_V$ from Table \ref{mdottable} to the presumed accretion luminosity $M_\mathrm{bol}(M_1,M_2)$ due to gravitational-wave radiation via
\beq
M_V = M_\mathrm{bol}(M_1,M_2) - \mathrm{BC} + \Delta M_V(i)
\eeq
where
\beq
\Delta M_V(i) = -2.5\log\left(\frac{\cos i}{0.5}\right)
\eeq
is the correction due to the observer observing a different area of the disc than the direction-averaged value, $\langle \cos i\rangle = 0.5$ (the unknown effect of limb darkening is not taken into account).
For HP Lib and V803 Cen the inclination is constrained by spectroscopic measurements of the velocity amplitude of the `bright spot' (the presumed accretion stream--accretion disk impact region) in these systems (\citealt{Roe06c}; see also section \ref{qV803Cen}). We make the assumption that the radial velocity semi-amplitude of the donor star is equal to that of the bright spot, which is usually a good estimate. Similar data does not exist for CR Boo and we thus adopt the value of $i=30^\circ$ derived from spectral line modeling by \citet{Nas01}, with the sidenote that their model-derived inclinations for AM CVn, HP Lib, and V803 Cen are in excellent agreement with the kinematically derived values. Table~\ref{mdottable} lists the inclinations we employ for all systems.

With the above assumptions we have all the ingredients needed to solve for the masses and mass transfer rates of our AM CVn stars: for a given mass ratio $q$, there is only one set of $M_1$ and $M_2$ for which the resulting accretion luminosity driven by gravitational-wave emission matches with the observed luminosity. For AM CVn, which has a kinematically measured mass ratio, the dominant contributions to the error on the mass transfer rate are the error on its absolute magnitude $M_V$ and the uncertainty in the bolometric correction. For HP Lib, CR Boo, and V803 Cen, we consider the uncertainty on their true mass ratios to be the most important factor. The errors on the mass transfer rates for these systems are estimated by varying $q$ by $\pm 50\%$ from the `best estimates' given in Table \ref{mdottable}. The resulting lower limits then correspond to the low mass ratios employed by \citet{Del05}, while the resulting upper limits correspond to $q$s that are such that $q_\mathrm{HP\,Lib}$ coincides with our kinematically measured $q_\mathrm{AM\,CVn}$, which one might expect if they share a similar evolutionary history, given their similar orbital period. Note that these ranges in $q$ give ranges of $M_2$ and $M_1$ (see Table \ref{mdottable}) that are anti-correlated. For GP Com, as mentioned, we can put upper limits on the accretion rate by assuming that the accretion luminosity has to be less than the observed luminosity (the latter of which is probably due in part to the accreting white dwarf). This leads to upper limits on $M_{1,2}$ under the assumption that the mass transfer rate is set by the rate of gravitational-wave emission. The lower limits on $M_{1,2}$ correspond to the minimum mass a Roche-lobe-filling helium object can have at the observed orbital period of GP Com.

We can compare our values to those obtained by \citet{Nas01} from spectral modeling of the accretion disks in these systems. With our larger distance for AM CVn, the accretion rate based on their models should be a bit larger than our value, at $\dot M_\mathrm{AM\,CVn}\sim 1\times 10^{-8}\,M_\odot$/yr (J.-E. Solheim, private communication). For the other systems they find $\dot M\sim 4\times 10^{-9}\,M_\odot$/yr, which is again larger than our values of $\dot M\sim 1\times 10^{-9}\,M_\odot$/yr. \citet{Elk00} on the other hand find $\dot M_\mathrm{AM\,CVn}\sim 2\times 10^{-9}\,M_\odot$/yr, also from spectral modeling of the accretion disk, which is lower than our value. Their value of $\dot M_\mathrm{CR\,Boo}\sim 1\times 10^{-9}\,M_\odot$/yr is close to ours.

\begin{deluxetable*}{rccccc}
\tablecaption{Apparent Mass Accretion Rates, Inferred Masses, and Gravitational-Wave Strain Amplitudes\label{mdottable}} 
\tablewidth{0pt}
\tablehead{\colhead{Parameter} & \colhead{AM CVn} &\colhead{HP Lib}&\colhead{CR Boo}&\colhead{V803 Cen}&\colhead{GP Com}}
\startdata
$P_{\rm orb}$ (s)       &1029                &1103               &1471               &1596              &2794\\
$q$                     &0.18\tnm{a}         &$0.06-0.18$\tnm{b} &$0.04-0.13$\tnm{b} &$0.05-0.14$\tnm{b}&0.018\tnm{a}\\
$i$ (degrees)           &$43\pm2$            &$26-34$            &$30$\tnm{c}        &$12-15$           &n/a\\
BC                      &$-2.5\pm0.3$        &$-2.5\pm0.3$       &$-2.5\pm0.3$       &$-2.5\pm0.3$      &$-0.5\pm0.2$\\[0.7ex]

$\dot{M}$ (M$_\odot$ yr$^{-1}$)  &\gpm{7.1}{2.2}{1.5}$\times 10^{-9}$  &$[0.81-2.2]\times 10^{-9}$  &$[0.38-1.2]\times 10^{-9}$  &$[0.57-1.6]\times 10^{-9}$  &$<3.6\pm0.5\times 10^{-12}$\\[0.7ex]

$M_2$ (M$_\odot$)       &$0.13\pm0.01$       &$0.048-0.088$      &$0.044-0.088$      &$0.059-0.109$     &$0.009-0.012$\\
$M_1$ (M$_\odot$)       &$0.71\pm0.07$       &$0.80-0.49$        &$1.10-0.67$        &$1.17-0.78$       &$0.50-0.68$\\[0.7ex]

$L$ (W)\tnm{d}          &\gpm{2.2}{1.4}{0.8}$\times 10^{27}$   &\gpm{3.4}{1.3}{0.9}$\times 10^{26}$   &\gpm{3.5}{1.2}{1.7}$\times 10^{26}$   &\gpm{6.5}{2.1}{3.1}$\times 10^{26}$   &$<$\gpm{1.1}{0.3}{0.2}$\times 10^{24}$\\[0.7ex]
$h$			&\gpm{2.0}{0.4}{0.3}$\times 10^{-22}$  &\gpm{3.7}{0.6}{0.8}$\times 10^{-22}$  &\gpm{2.1}{0.4}{0.5}$\times 10^{-22}$  &\gpm{3.0}{0.5}{0.7}$\times 10^{-22}$  &$[4.0-6.6]\times 10^{-23}$\\
\enddata
\tablenotetext{a}{Measured from kinematics.}
\tablenotetext{b}{Inferred from superhump period; note the differences with Table~1 of \citet{Del05}.}
\tablenotetext{c}{From \citet{Nas01}.}
\tablenotetext{d}{Inferred accretion luminosity, Eq.\ (\ref{Lbol}).}
\end{deluxetable*}

\subsection{The Nature of the Donor Stars}
The most surprising result of our \emph{HST} parallaxes is the large distance to AM CVn, which exceeds every previous estimate. The resulting large luminosity of AM CVn already suggests a relatively high mass accretion rate. In addition to the large distance found here, AM CVn's recently measured mass ratio, too, was found to be significantly larger than previously thought \citep{Roe06b}. With these two pieces of information we could solve self-consistently for the masses of the components (within the assumptions outlined in Section \ref{accsec}), and provide strong evidence that the donor star in AM CVn is a relatively massive, semi-degenerate helium object.

\begin{figure}[!th]
\begin{center}
\plotone{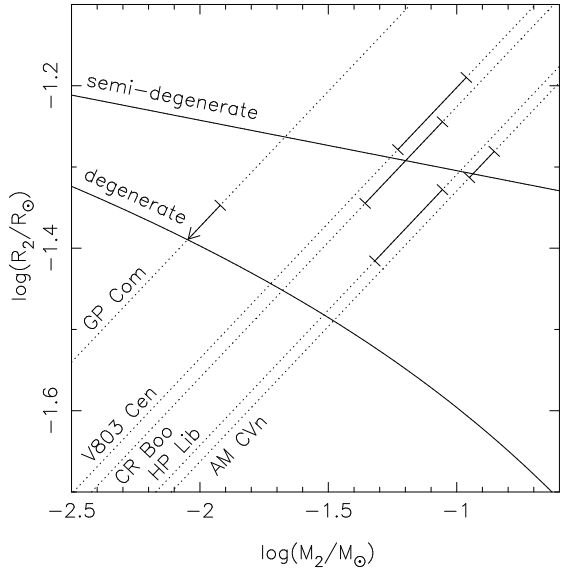}
\caption{Constraints on the masses and radii of the donor stars, compared to degenerate and semi-degenerate evolutionary tracks from \citet{Nel01}. The dotted diagonals represent the Roche-lobe filling solutions for the different systems. The solid-line regions represent the constraints on the masses from Table \ref{mdottable}. We see that, apart from GP Com, the donors appear to be rather semi-degenerate.}
\label{massesradii}
\end{center}
\end{figure}

\begin{figure}[!h]
\begin{center}
\plotone{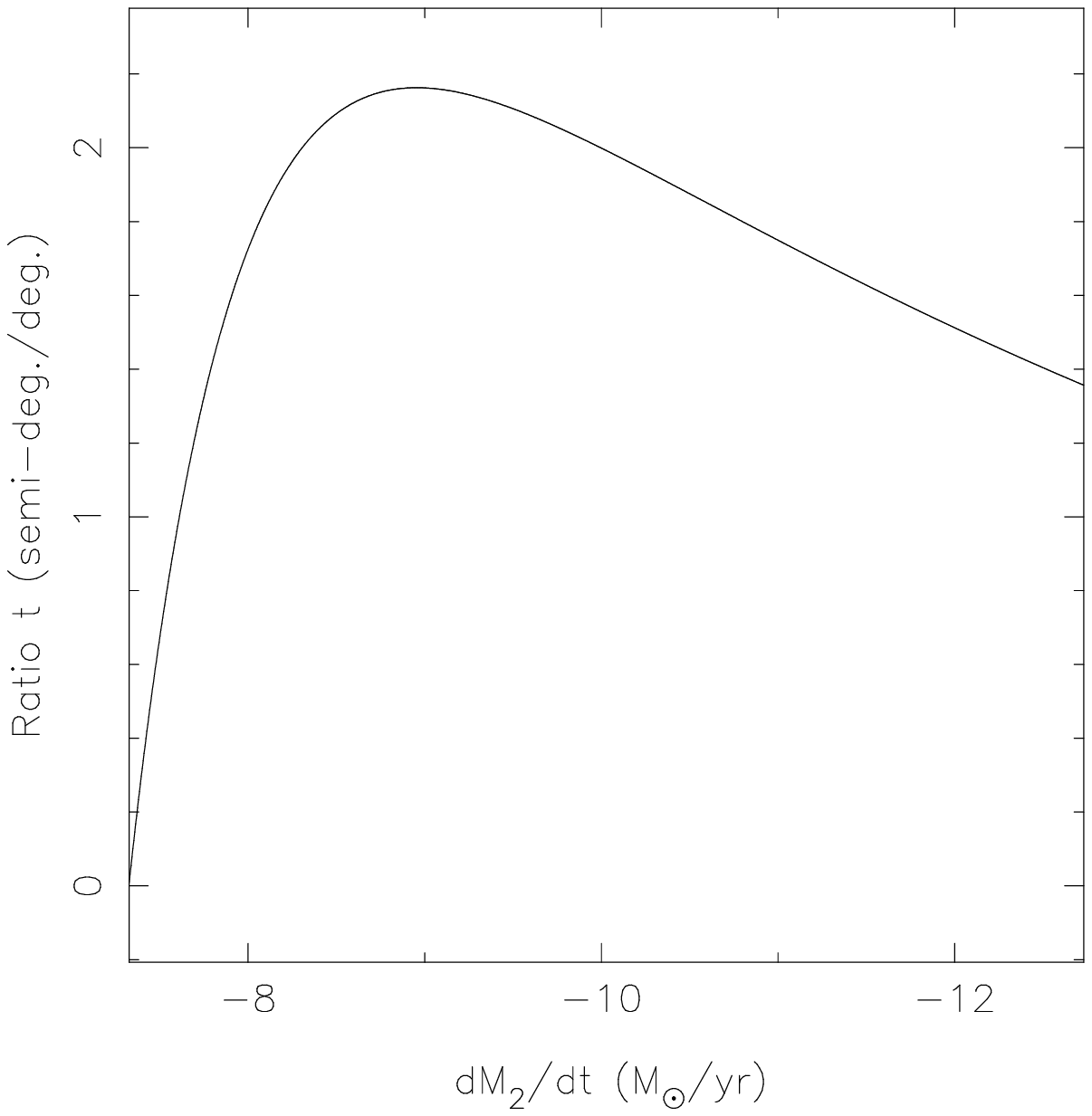}
\caption{Ratio of the integrated times spent at a given mass transfer rate \emph{or higher}, for a $0.6\,M_\odot$ white dwarf accreting from a fully degenerate ($0.25\,M_\odot$) and a semi-degenerate ($0.2\,M_\odot$) donor star as modeled by \citet{Nel01}. Semi-degenerate systems are seen to spend about twice as long at phases of relatively high mass transfer, $10^{-10}\lesssim\dot{M_2}\lesssim10^{-8}\,M_\odot$/yr, corresponding to the mass transfer rates found for AM CVn, HP Lib, CR Boo and V803 Cen.}
\label{selection}
\end{center}
\end{figure}

For HP Lib, CR Boo, and V803 Cen, there is a larger uncertainty in the masses of the components due to the larger uncertainties on their mass ratios, as discussed in the previous section. Nevertheless, if we compare the data with evolutionary predictions for fully-degenerate and semi-degenerate donors (see Figure \ref{massesradii}), the data are seen to favor rather massive, semi-degenerate donor stars in all systems but GP Com. This agrees qualitatively with the accretion disk models of \citet{Nas01}, but disagrees with \citet{Elk00} who find a fully-degenerate donor in AM CVn.

It should be noted that there is a selection effect against fully-degenerate donor stars, since AM CVn stars with semi-degenerate donors are expected to be brighter at short orbital periods due to the higher mass accretion rates. At long orbital periods this is partially offset by the fact that semi-degenerate systems evolve to lower accretion rates more quickly. If we evolve a fully-degenerate and a semi-degenerate system, based on the parameterization of \citet{Nel01}, we see that semi-degenerate donors are in a bright state about twice as long, where we take `bright state' to be $\dot{M_2}\sim10^{-8}-10^{-10}\,M_\odot$/yr, i.e.\ a state corresponding to the four bright AM CVns under consideration here. See Figure \ref{selection}.

We thus observe four relatively massive, semi-degenerate systems out of five, with a potential factor-of-two selection effect in favor of the semi-degenerates. We can conclude that at least a significant fraction, and possibly all, of the short-period systems (at $P_\mathrm{orb}\lesssim2000$\,s) have hot, semi-degenerate donors rather than cold, degenerate ones. AM CVn stars with such hot, semi-degenerate donors are expected naturally from the helium-star channel \citep{Nel01}, but a fraction of AM CVn progenitors in the WD channel are also expected to still have hot donors upon coming into contact \citep{Del05}. Conceivably, it could be predominantly this hot fraction of systems that survive as AM CVn stars in the WD channel, since the mass transfer rate upon Roche-lobe overflow can be up to two orders of magnitude lower than for cold, degenerate donors \citep{Nel01,Del05,Del07}. Another possibility would be that cold white dwarf donors in the WD channel are reheated at some stage, for instance by tidal heating close to orbital period minimum (e.g.\ \citealt{Ibe98}) or in a thermonuclear event such as those proposed by \citet{Bil07}. Interpretation of our results in terms of the evolutionary history of the systems requires a more detailed study of these effects. Of particular interest would be a comparison between HP Lib and AM CVn, whose donor stars appear to have a different level of degeneracy despite their otherwise identical appearance \citep{Roe06b,Roe06c}. A kinematic measurement of HP Lib's mass ratio would be needed to determine whether its donor star is truly more degenerate.

We remark that there is one, but really only one, crucial assumption in our analysis that leads us to conclude that the donor stars in AM CVns are semi-degenerate objects, namely, that the mass transfer rate is set by the emission of gravitational waves. If there should be an additional, dominant driver of mass transfer, we could be overestimating the masses enough (based on the observed luminosities) that the donors could in reality be fully degenerate. Within the assumption of gravitational-wave-driven mass transfer, the result that the donors are semi-degenerate is quite robust. This is a consequence of the accretion luminosity due to gravitational waves being strongly dependent on the masses of the stars, as can readily be judged from Eqs.\ (\ref{Lbol})--(\ref{Jdot}). The uncertainty in the inferred masses is therefore small compared to the uncertainties in the bolometric luminosities of the systems.

The possibility of the mass transfer rate (due to gravitational waves) being out of equilibrium cannot be ruled out, but we do remark that it is quite difficult to change the donor star's radius by for instance irradiation, since the structure of the donor star is largely set by degeneracy pressure (C. Deloye, private communication). At late times ($P_\mathrm{orb}\gtrsim 45$\,min) a hot, semi-degenerate donor may start to cool and contract towards a fully-degenerate state, which could cause the mass transfer rate to go down temporarily \citep{Del07}. This may be relevant for GP Com.

\subsection{The Mass Ratio of V803 Cen}
\label{qV803Cen}
The small mass ratio of V803 Cen, $q_\mathrm{V803\,Cen}=0.016$, implied by the small superhump period excess as obtained from \citet{Pat01} by \citet{Del05} and others, leads to a necessarily large accretor mass close to the Chandrasekhar limit, and a small donor mass close to the minimum mass for a Roche-lobe filling secondary (corresponding to a cold, fully-degenerate object). This was already noted by \citet{Del05}. Given these implications, it should be noted that V803 Cen is the one system for which the photometric superhump period excess appeared to be uncertain due to an uncertain identification of the 1611-second orbital period (see also \citealt{Pat00}), which has nevertheless entered the literature (\citealt{Pat01}, \citealt{Pat02}, \citealt{Wou03}, \citealt{Del05}).

We have recently measured a significantly shorter orbital period, $P_\mathrm{V803\,Cen}=1596.4\pm1.2$ seconds, from VLT spectroscopy \citep{Roe06c}. The mass ratio we use in this paper is the one we obtain from the corresponding (larger) superhump period excess, via relation (\ref{epsilon}), again with a $\pm50\%$ range. Since it is not completely clear whether the 1618\,s superhump period or the 1611\,s period (both from \citealt{Pat00}) is the more appropriate one to use for calculating the superhump period excess, we have used the average of the two.

The resulting mass ratio $q_\mathrm{V803\,Cen}=0.05-0.14$ lifts the strong constraints on the mass of the donor (cf.\ \citealt{Del05}). A semi-degenerate donor is then strongly favored, as shown in Section \ref{accsec}, and the accretor does not have to be very close to the Chandrasekhar mass in order to explain the observed luminosity of the system.

\subsection{AM CVn Stars as Sources of Gravitational Waves}
\label{gwrsec}
The ultra-compact nature of the AM CVn stars, combined with the relative proximity of the known systems, makes them the strongest known sources of gravitational waves in the frequency regime covered by space-borne gravitational-wave detectors such as \emph{LISA}.

\begin{figure}[!t]
\begin{center}
\plotone{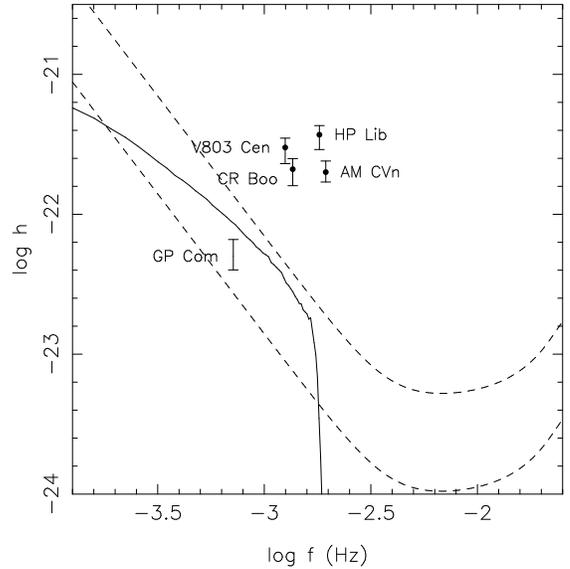}
\caption{Predicted gravitational-wave strain amplitudes $h$ and frequencies $f$ of the five AM CVn stars for which we obtained an \emph{HST} parallax. The upper and lower dashed lines show the design sensitivities of \emph{LISA} for a signal-to-noise ratio of 5 and 1, respectively, in one year of data-collecting \citep{Lar05}. The solid line is a population synthesis prediction for the confusion-limited Galactic background, also for a mission duration of one year \citep{Nel04}.}
\label{gwrfig}
\end{center}
\vspace{14ex}
\end{figure}

The masses of the stars, inferred from their luminosities, can be used to estimate the gravitational-wave strain amplitudes of our five AM CVns. The gravitational-wave strain amplitude $h$ is given by (see \citealt{Roe06b}, after \citealt{Tim06})
\begin{eqnarray}
h &=& 2.84 \cdot 10^{-22}\sqrt{\cos^4i+ 6\cos^2i + 1}\nonumber\\
&&\phantom{=}\times\left(\frac{\mathcal{M}}{M_\odot}\right)^{5/3} \left(\frac{P_\mathrm{orb}}{\mathrm{1\,hr}}\right)^{-2/3} \left(\frac{d}{\mathrm{1\,kpc}}\right)^{-1}
\end{eqnarray}
where $\mathcal{M}=(M_1M_2)^{3/5}/(M_1+M_2)^{1/5}$ is the so-called chirp mass, and $d$ the distance to the star.
Figure \ref{gwrfig} shows the resulting strain amplitudes of our five systems. These values are quite insensitive to the uncertain mass ratios $q$ of some of our systems; their errors are dominated by the errors on their bolometric luminosities. We see that the four short-period systems are all excellent test sources for \emph{LISA}, standing out significantly above both the instrument's design sensitivity and the expected average Galactic background signal (although a more rigorous analysis may be needed to predict their exact signal-to-noise ratios; see \citealt{Str06}). The estimates for the four bright systems have gone up from initial estimates by \citet{Nel05} due to their rather massive donors, and their relatively low inclinations. The expectations for GP Com remain rather low: it does not stand out from the instrumental and Galactic noise and will be much more difficult to identify. We note that, as long as the uncertainty about the orbital frequency is larger than the frequency resolving power of \emph{LISA} ($\Delta f \sim t_\mathrm{obs}^{-1}\sim 10^{-8}$\,Hz, for a mission duration $t_\mathrm{obs} = 1$\,yr), the chances of confusion with Galactic noise sources decrease greatly with more accurately known orbital periods for these systems. The orbital period of AM CVn is known to an accuracy of better than one part per million from large time-base photometric monitoring ($P_\mathrm{AM\,CVn}=1028.7322\pm0.0003$\,s, \citealt{Ski99}), and the orbital period of HP Lib also seems quite secure ($P_\mathrm{HP\,Lib}=1102.70\pm0.05$\,s, \citealt{Pat02}), but the orbital periods of especially V803 Cen and CR Boo seem to be less well-known. It would be worthwhile to measure the orbital periods of these systems with greater accuracy.

\acknowledgments
GHAR, PJG and LMR are supported by NWO-VIDI grant 639.042.201
to P.J. Groot. GN was supported by NWO-VENI grant 639.041.405 to
G. Nelemans. TRM was supported by a PPARC Senior 
Research Fellowship. DS acknowledges a Smithsonian Astrophysical Observatory Clay Fellowship.
Support for this work was provided by NASA through grants GO-09168 and
GO-09348 from the Space Telescope Science Institute, which is operated
by AURA, Inc., under NASA contract NAS~5-26555.
These results are based partially on observations made with the
Isaac Newton Telescope and the William Herschel Telescope,
operated on the island of La Palma by the Isaac Newton Group
in the Spanish Observatorio del Roque de los Muchachos of the
Instituto de Astrof\'isica de Canarias, and additional observations
obtained with the Magellan--Baade telescope at Las Campanas Observatory, Chile.
We thank Jan-Erik Solheim and Chris Deloye for stimulating
discussion during the preparation of this paper, and Ken Shen for kindly providing white dwarf model data.

\clearpage
\appendix
\section{AM CVn and Reference Star Data Tables}
\clearpage

\begin{deluxetable}{l l r r}
\tablewidth{0in}
\tablecaption{Log of Observations and FGS Photometry\label{tbl-LOO}}
\tablehead{\colhead{Set}&\colhead{MJD}&\colhead{Roll (\arcdeg)\tablenotemark{a}}&\colhead{$V$\tablenotemark{b}}}
\startdata
AM CVn-1&52222.81198&217&14.02$\pm$0.01\\
AM CVn-2&52225.88485&58&14.01$\pm$0.01\\
AM CVn-3&52387.43705&217&14.04$\pm$0.02\\
AM CVn-4&52390.64253&33&14.03$\pm$0.01\\
AM CVn-5&52431.24828&33&14.03$\pm$0.01\\
AM CVn-6&52435.52559&58&13.99$\pm$0.01\\
AM CVn-7&52715.78999&58&14.01$\pm$0.01\\
AM CVn-8&52719.72357&359&14.04$\pm$0.02\\
AM CVn-9&52796.35781&359&14.04$\pm$0.02\\
AM CVn-10&52803.28106&58&14.03$\pm$0.02\\
\\
CR Boo-1&52095.17132&82&14.51$\pm$0.03\\
CR Boo-2&52103.46567&82&14.16$\pm$0.01\\
CR Boo-3&52136.09742&83&14.36$\pm$0.01\\
CR Boo-4&\ldots&\ldots&\ldots\\
CR Boo-5&52309.56485&273&13.93$\pm$0.03\\
CR Boo-6&52322.84773&273&15.99$\pm$0.04\\
CR Boo-7&52501.29821&83&15.88$\pm$0.04\\
CR Boo-8&52512.98296&87&15.96$\pm$0.02\\
CR Boo-9&52676.15807&273&15.61$\pm$0.04\\
CR Boo-10&52686.69897&273&15.07$\pm$0.01\\
\\
V803 Cen-1&52112.02035&68&14.19$\pm$0.03\\
V803 Cen-2&52115.43291&68&13.50$\pm$0.05\\
V803 Cen-3&52161.84252&37&13.64$\pm$0.01\\
V803 Cen-4&52164.04853&37&15.68$\pm$0.08\\
V803 Cen-5&52294.81707&247&13.46$\pm$0.01\\
V803 Cen-6&52300.82760&247&13.51$\pm$0.01\\
V803 Cen-7&52526.14146&37&14.64$\pm$0.03\\
V803 Cen-8&52530.14615&37&13.85$\pm$0.04\\
V803 Cen-9&52659.74478&246&14.34$\pm$0.02\\
V803 Cen-10&52664.41432&246&13.69$\pm$0.04\\
\\
HP Lib-1&52135.63781&74&13.61$\pm$0.03\\
HP Lib-2&52138.24565&74&13.62$\pm$0.01\\
HP Lib-3&52168.26915&80&13.59$\pm$0.02\\
HP Lib-4&52172.41573&80&13.59$\pm$0.02\\
HP Lib-5&52318.71790&256&13.60$\pm$0.02\\
HP Lib-6&52326.79246&256&13.60$\pm$0.02\\
HP Lib-7&52532.15146&80&13.57$\pm$0.02\\
HP Lib-8&52539.16093&80&13.58$\pm$0.02\\
HP Lib-9&52680.90353&256&13.56$\pm$0.02\\
HP Lib-10&52693.71125&256&13.54$\pm$0.02\\
\\
GP Com-1&52088.34598&65&15.96$\pm$0.01\\
GP Com-2&52097.16817&65&15.99$\pm$0.01\\
GP Com-3&52234.70052&229&15.89$\pm$0.02\\
GP Com-4&52242.04902&229&15.92$\pm$0.01\\
GP Com-5&52270.97565&245&15.93$\pm$0.01\\
GP Com-6&52273.91306&245&15.94$\pm$0.02\\
GP Com-7&52598.75738&229&15.93$\pm$0.02\\
GP Com-8&52633.85590&244&16.01$\pm$0.02\\
GP Com-9&52637.92521&244&16.00$\pm$0.02\\
GP Com-10&52963.80643&229&15.89$\pm$0.01\\
\enddata
\tablenotetext{a}{Spacecraft roll as defined in Chapter 2, FGS Instrument Handbook \citep{Nel02}.}
\tablenotetext{b}{Average of 2 to 4 observations at each epoch. Errors are internal and are the standard deviation at each epoch.}
\end{deluxetable}

\begin{deluxetable}{lrr}
\tablewidth{0in}
\tablecaption{Reference Star Positions\label{tbl-POS}}
\tablehead{\colhead{ID}&
\colhead{$\xi$ \tablenotemark{a}} &
\colhead{$\eta$ \tablenotemark{a}} 
}
\startdata
AM CVn\tablenotemark{b}&&\\
ref-2&$-$77.5390 0.0003&$-$137.4972 0.0002\\
ref-3&159.3017 0.0003&$-$173.8572 0.0002\\
ref-5&60.260 0.0003&$-$88.9279 0.0002\\
CR Boo\tablenotemark{c}&&\\
ref-2&$-$192.2319 0.0003&$-$49.5013 0.0002\\
ref-3&$-$242.6593 0.0002&$-$3.4712 0.0001\\
ref-4&304.3653 0.0003&-16.0044 0.0002\\
ref-5&250.2984 0.0006&19.8065 0.0004\\
ref-6&437.8449 0.0009&9.7262 0.0006\\
V803 Cen\tablenotemark{d}&&\\
ref-2&78.2233 0.0003&42.8896 0.0002\\
ref-3&$-$119.9175 0.0003&$-$26.1711 0.0002\\
ref-4&$-$55.0075 0.0002&110.9745 0.0002\\
ref-5&$-$122.0958 0.0002&67.3278 0.0002\\
ref-6&137.4346 0.0002&39.0072 0.0002\\
HP Lib\tablenotemark{e}&&\\
ref-2&89.3402 0.0003&$-$66.1514 0.0002\\
ref-3&249.1257 0.0003&$-$18.3677 0.0002\\
ref-4&9.8453 0.0004&77.8612 0.0003\\
ref-5&$-$4.4333 0.0003&55.4678 0.0002\\
ref-6&$-$311.6452 0.0003&$-$13.1930 0.0002\\
GP Com\tablenotemark{f}&&\\
ref-2&81.0209 0.0002&33.9032 0.0001\\
ref-3&88.3580 0.0002&95.1984 0.0001\\
ref-4&80.9519 0.0002&$-$73.6567 0.0001\\
ref-5&105.5746 0.0003&30.4648 0.0002\\
ref-6&$-$103.6979 0.0002&101.5386 0.0002\\
\enddata
\tablenotetext{a}{\,$\xi$ and $\eta$ are relative RA and Dec in arcseconds}
\tablenotetext{b}{\,$12^\mathrm{h}34^\mathrm{m}54\fs58$, $+37^\circ37'43\farcs4$, J2000, epoch 2002.44}
\tablenotetext{c}{\,$13^\mathrm{h}48^\mathrm{m}55\fs29$, $+07^\circ57'34\farcs8$, J2000, epoch 2002.14}
\tablenotetext{d}{\,$13^\mathrm{h}23^\mathrm{m}44\fs51$, $-41^\circ44'30\farcs4$, J2000, epoch 2002.07}
\tablenotetext{e}{\,$15^\mathrm{h}35^\mathrm{m}53\fs08$, $-14^\circ13'12\farcs3$, J2000, epoch 2002.14}
\tablenotetext{f}{\,$13^\mathrm{h}05^\mathrm{m}42\fs85$, $+18^\circ01'02\farcs6$, J2000, epoch 2001.99}
\end{deluxetable}

\begin{deluxetable}{lrrrrrr}
\tablewidth{0in}
\tablecaption{Reference Star Spectrophotometric Parallaxes\label{tbl-SPP}}
\tablehead{
\colhead{ID}& \colhead{Spectral~Type}&
\colhead{$V$} & \colhead{$M_V$} & \colhead{$A_V$} &\colhead{$m-M$}& 
\colhead{$\pi_\mathrm{abs}$(mas)}
} 
\startdata
AM CVn&&&&&&\\
ref-2   &G4V    &14.37  &5.0    &0.05   &9.4$\pm$0.7    &1.3$\pm$0.4\\
ref-3   &K6V    &15.15  &7.6    &0.05   &7.5 0.7        &3.2 1.0\\
ref-5   &G4V    &12.42  &5.0    &0.05   &7.5 0.7        &3.3 1.0\\
CR Boo  &       &       &       &       &               &\\
ref-2   &G9V    &15.57  &5.74   &0.02   &9.8 0.7        &1.1 0.4\\
ref-3   &M2III  &11.40  &$-$0.6 &0.02   &12.0 0.7         &0.4 0.1\\
ref-4\tablenotemark{a}&F2V    &6.81    &3.0    &0.02   &3.8 0.7        &16.9 0.9\\
ref-5   &G0V    &14.82  &4.4    &0.02   &10.4 0.7       &0.8 0.3\\
ref-6   &G7V    &11.93  &5.4    &0.02   &6.5 0.7        &5.1 1.6\\
V803 Cen&       &       &       &       &&\\
ref-2   &K1V    &15.86  &6.2    &0.31   &9.7 0.7        &1.3 0.4\\
ref-3   &G4V    &15.34  &5.0    &0.31   &10.4 0.7       &0.9 0.3\\
ref-4   &F6.5V  &13.38  &3.8    &0.31   &9.6 0.7        &1.4 0.4\\
ref-5   &G6V    &14.17  &5.3    &0.31   &8.9 0.7        &1.9 0.6\\
ref-6   &F8V    &14.15  &4.0    &0.31   &10.2 0.7       &1.1 0.4\\
HP Lib  &       &       &       &       &&\\
ref-2   &F5V    &12.86  &3.5    &0.34   &9.4 0.7        &1.6 0.5\\
ref-3   &G9V    &14.06  &5.7    &0.34   &8.4 0.7        &2.6 0.8\\
ref-4   &K7V    &12.57  &7.9    &0.34   &4.7 0.7        &14.0 4.5\\
ref-5   &K4V    &11.80  &7.1    &0.34   &4.7 0.7        &13.6 4.4\\
ref-6   &A2.5V  &14.60  &1.3    &0.34   &13.3 1.0       &0.3 0.1\\
GP Com  &       &       &       &       &&\\
ref-2   &G0V    &15.17  &4.4    &0.02   &10.8 0.7       & 0.7 0.2\\
ref-3   &G8V    &14.39  &5.58   &0.02   &8.8 0.7        &1.7 0.6\\
ref-4   &K0V    &14.25  &5.58   &0.02   &8.7 0.7        &1.8 0.6\\
ref-5   &K5V    &13.56  &7.4    &0.02   &6.2 0.7        &5.7 1.9\\
ref-6   &G8V    &15.12  &5.58   &0.02   &9.5 0.7        &1.2 0.4\\
\enddata 
\tablenotetext{a}{= HIP 67379}
\end{deluxetable}

\begin{deluxetable}{lrrrrr}
\tablewidth{0in}
\tablecaption{FGS and Near-IR Reference Star Photometry\label{tbl-IR}}
\tablehead{\colhead{ID}&
\colhead{$V$} &
\colhead{$K$} &
\colhead{$J-H$} &
\colhead{$J-K$} &
\colhead{$V-K$}
}
\startdata
AM CVn&&&&&\\
ref-2&14.37 0.1&12.79 0.03&0.44 0.04&0.43 0.03&1.58 0.10\\
ref-3&15.15 0.1&12.09 0.02&0.74 0.04&0.82 0.03&3.06 0.10\\
ref-5&12.42 0.1&10.88 0.02&0.41 0.04&0.42 0.03&1.54 0.10\\
CR Boo&&&&&\\
ref-2&15.57 0.1&13.64 0.05&0.47 0.05&0.48 0.05&1.93 0.11\\
ref-3&11.4 0.1&7.13 0.05&0.91 0.02&1.13 0.05&4.28 0.11\\
ref-4&14.8 0.1&5.93 0.01&0.21 0.04&0.22 0.02&0.87 0.05\\
ref-5&6.8 0.1&13.54 0.05&0.34 0.05&0.27 0.06&1.46 0.11\\
ref-6&11.89 0.1&\ldots&\ldots&\ldots&\ldots\\
V803 Cen&&&&&\\
ref-2&15.86 0.1&13.42 0.04&0.53 0.04&0.62 0.04&2.44 0.11\\
ref-3&15.34 0.1&13.58 0.04&0.45 0.04&0.43 0.05&1.76 0.11\\
ref-4&13.38 0.1&11.96 0.02&0.29 0.03&0.33 0.03&1.42 0.10\\
ref-5&14.17 0.1&12.29 0.02&0.42 0.03&0.51 0.03&1.88 0.10\\
ref-6&14.15 0.1&12.45 0.03&0.35 0.03&0.36 0.04&1.70 0.10\\
HP Lib&&&&&\\
ref-2&12.88 0.1&11.40 0.02&0.30 0.03&0.34 0.03&1.48 0.10\\
ref-3&14.11 0.1&11.93 0.02&0.51 0.03&0.51 0.03&2.18 0.10\\
ref-4&12.6 0.1&9.17 0.02&0.69 0.03&0.84 0.03&3.43 0.10\\
ref-5&11.84 0.1&8.87 0.02&0.66 0.03&0.77 0.03&2.97 0.10\\
ref-6&14.6 0.1&\ldots&\ldots&\ldots&\ldots\\
GP Com&&&&&\\
ref-2&15.17 0.1&13.73 0.05&0.34 0.04&0.36 0.06&1.44 0.11\\
ref-3&14.39 0.1&12.62 0.03&0.43 0.03&0.46 0.03&1.77 0.10\\
ref-4&14.25 0.1&12.24 0.03&0.46 0.03&0.54 0.03&2.01 0.10\\
ref-5&13.56 0.1&10.71 0.02&0.67 0.02&0.74 0.03&2.85 0.10\\
ref-6&15.12 0.1&13.44 0.04&0.40 0.04&0.53 0.05&1.68 0.11\\
\enddata
\end{deluxetable}

\begin{deluxetable}{lrrrrrr}
\tablewidth{0in}
\tablecaption{Field $A_V$ from Reference Star Spectrophotometry\label{tbl-AV}}
\tablehead{\colhead{ID}&
\colhead{Spectral~Type}&   \colhead{$(V-K)_0$}&  \colhead{$V-K$} &  \colhead{$E(V-K)$}&
\colhead{$A_V$\tablenotemark{a}}&\colhead{$\langle A_V\rangle$}
}
\startdata
AM CVn&&&&&&0.05$\pm$0.02\\
ref-2&G4V&1.52&1.58&0.06&0.07&\\
ref-3&K6V&3.01&3.06&0.05&0.06&\\
ref-5&G4V&1.52&1.54&0.02&0.02&\\
CR Boo&&&&&&0.03$\pm$0.04\\
ref-2&G9V&1.88&1.93&0.05&0.05&\\
ref-3&M2III&4.30&4.28&$-$0.02&$-$0.02&\\
ref-4&F2V&0.86&0.87&0.01&0.01&\\
ref-5&G0V&1.41&1.46&0.05&0.06&\\
V803 Cen&&&&&&0.31$\pm$0.09\\
ref-2&K1V&2.14&2.44&0.30&0.33&\\
ref-3&G4V&1.52&1.76&0.24&0.26&\\
ref-4&F6.5V&1.19&1.42&0.23&0.26&\\
ref-5&G6V&1.67&1.88&0.21&0.23&\\
ref-6&F8V&1.28&1.70&0.42&0.46&\\
HP Lib&&&&&&0.34$\pm$0.06\\
ref-2&F5V&1.10&1.48&0.38&0.42&\\
ref-3&G9V&1.88&2.18&0.3&0.33&\\
ref-4&K7V&3.17&3.43&0.26&0.29&\\
ref-5&K4V&2.68&2.97&0.29&0.32&\\
GP Com&&&&&&0.02$\pm$0.11\\
ref-2&G0V&1.41&1.44&0.03&0.03&\\
ref-3&G8V&1.80&1.77&$-$0.03&$-$0.03&\\
ref-4&K0V&1.96&2.01&0.05&0.05&\\
ref-5&K5V&2.70&2.85&0.15&0.16&\\
ref-6&G8V&1.80&1.68&$-$0.12&$-$0.14&\\
\enddata
\tablenotetext{a}{$A_V = 1.1E(V-K)$}
\end{deluxetable}

\end{document}